\documentclass[fleqn,usenatbib]{mnras}

\usepackage{newtxtext,newtxmath}

\usepackage[T1]{fontenc}

\DeclareRobustCommand{\VAN}[3]{#2}
\let\VANthebibliography\thebibliography
\def\thebibliography{\DeclareRobustCommand{\VAN}[3]{##3}\VANthebibliography}

\usepackage{graphicx}	
\usepackage{amsmath}	
\usepackage{booktabs}	
\usepackage{url}		

\newcommand{\hMpc}{\,h^{-1}\,\mathrm{Mpc}}	
\newcommand{\hkpc}{\,h^{-1}\,\mathrm{kpc}}	
\newcommand{\nel}{n_{\mathrm{e}}}			
\newcommand{\rhodm}{\rho_{\mathrm{DM}}}	
\newcommand{\dex}{\,\mathrm{dex}}

\title[CosmoMap: hot gas from dark matter]{CosmoMap: Predicting Hot Gas Temperature and Density from Dark Matter with a Conditioned 3D U-Net}

\author[A. Bhutani and M. Lieu]{
Aman Bhutani,$^{1}$
Maggie Lieu$^{1}$
\\
$^{1}$School of Physics and Astronomy, University of Nottingham, University Park, Nottingham NG7 2RD, UK
}

\date{Accepted XXX. Received YYY; in original form ZZZ}

\pubyear{\the\year{}}

\begin{document}
\label{firstpage}
\pagerange{\pageref{firstpage}--\pageref{lastpage}}
\maketitle

\begin{abstract}
Mock X-ray observations of hot gas in galaxy groups, clusters and the cosmic web require its temperature and density over large volumes and many cosmologies, for which hydrodynamical simulations are expensive. We present CosmoMap, a 3D U-Net FiLM-conditioned on cosmological and feedback parameters that predicts gas temperature $T$ and electron density $\nel$ from dark matter density and velocity. On the CAMELS IllustrisTNG Latin hypercube, CosmoMap reaches $R^2 = 0.765$ for $\log T$ and $0.934$ for $\log\nel$ at $z=0$, reducing the error of gradient-boosted trees by 22--23 per cent. Temperature is predicted best in the densest regions, mostly galaxy- and group-scale haloes ($R^2 = 0.88$), but the electron density of the brightest X-ray-emitting gas is biased low by $0.12\dex$. One model outperforms the baselines at every redshift up to $z=2$, and using inputs from gravity-only simulations costs little accuracy. The predictions reproduce scaling relations, surface brightness profile shapes and most of the cosmological information in the gas. They produce box-integrated luminosities with a median bias of $-0.16\dex$, sensitive at the $0.2\dex$ level to the inference scheme unless batch normalisation is used. Training for per-voxel uncertainties with a heteroscedastic likelihood costs accuracy and lowers luminosities by a further $0.1$--$0.3\dex$; the composite-loss variant, CosmoMap-$\sigma$, provides uncertainties that are calibrated at 68 per cent after rescaling by predicted emissivity, except for the density of the brightest gas. Compact structures are smoothed: mock X-ray lightcones recover 89 per cent of the brightest tenth of sources but 28 per cent of all sources.
\end{abstract}




\section{Introduction}
\label{sec:intro}

The hot gas in galaxy groups, clusters and the densest filaments is a sensitive probe of both cosmology and baryonic physics \citep{Voit2005Tracing, Allen2011Cosmological, Kravtsov2012Formation}. At $10^6$--$10^8$\,K it radiates in X-rays mainly through thermal bremsstrahlung, with an emissivity scaling as $\nel^2 T^{1/2}$ \citep{Sarazin1986XrayEmission, Rybicki1985Radiative}. The first eROSITA all-sky survey catalogue already contains more than 12\,000 groups and clusters from half of the sky \citep{Bulbul2024eRASS1}, and the NewAthena mission \citep{Cruise2025NewAthena} will study fainter and more distant systems. Interpreting these surveys requires large numbers of mock observations to characterise selection effects and mass--observable relations, over the range of cosmologies and feedback models allowed by the data.

Hydrodynamical simulations such as IllustrisTNG \citep{Springel2018First, Pillepich2018First, Nelson2019IllustrisTNG} and SIMBA \citep{Dave2019SIMBA} model the gas self-consistently, but are far more expensive than gravity-only $N$-body simulations. Machine learning offers a way to bridge the two \citep{Dvorkin2022Machine}. Networks have been trained to predict galaxy properties \citep{Kasmanoff2020dm2gal, Zhang2019DM2Gal}, neutral hydrogen \citep{Wadekar2021HiNET} and thermal Sunyaev--Zel'dovich signals \citep{Thiele2020SZ} from dark matter, and several groups have ``painted'' hydrodynamical gas fields onto dark matter simulations with generative or Lagrangian models \citep{Troster2019Painting, Dai2021LDL, Horowitz2022HyPhy, Bernardini2022EMBER, Bernardini2025EMBER2}. Closest to our goal, \citet{Cerardi2025XXL} trained a Lagrangian deep learning emulator of the intracluster gas on CAMELS and used it to forward-model X-ray cluster counts for cosmological inference without explicit scaling relations, and \citet{Lee2026BIND} inpaint the baryonic fields of CAMELS groups and clusters onto gravity-only haloes with a conditional generative model. The CAMELS project \citep{VillaescusaNavarro2021CAMELS, VillaescusaNavarro2022Multifield} makes it possible to train such mappings across a systematically varied space of cosmological and feedback parameters. For X-ray forward modelling, such a mapping must predict both fields that set the emission, respond to the cosmology and feedback model, be accurate in the dense regions that dominate the signal, and be shown to outperform simpler alternatives.

In this paper we present CosmoMap, a 3D U-Net \citep{Ronneberger2015UNet} conditioned on the six CAMELS parameters through feature-wise linear modulation \citep[FiLM;][]{Perez2018FiLM}, which predicts gas temperature and electron density jointly from dark matter density and velocity. Relative to the approaches above, it combines joint three-dimensional prediction of the two fields that set the X-ray emission, explicit conditioning on all six cosmological and feedback parameters, and a benchmark against non-neural baselines and component ablations. We evaluate it under a single protocol, in which fields are normalised with fixed global statistics so that no property of a target field is used at test time, one simulation-level split is shared by every model, and training is repeated with several seeds. We test three things that matter for X-ray science:
\begin{enumerate}
\item field accuracy against simple and strong baselines and through an ablation study, which retrains the model with one component changed at a time, as a function of environment, scale, redshift and simulation parameters;
\item preservation of physical structure and X-ray emission properties: dark matter--gas scaling relations, the detection and surface brightness profiles of bright emission structures, X-ray luminosities, cosmological information, and predictive uncertainties;
\item mock X-ray lightcones with survey-like selection, compared source by source with lightcones built from the true fields.
\end{enumerate}

Two results shape the paper. Conditioning on the simulation parameters provides most of the network's advantage over simpler models, especially in low-density gas. And training with a heteroscedastic likelihood loss, a common way to obtain per-voxel uncertainties, costs accuracy and biases X-ray luminosities in each of the three forms we test, though to different degrees. We therefore train the main CosmoMap model with a mean-squared-error loss and use a likelihood-trained variant, CosmoMap-$\sigma$, for uncertainty estimates.

The CAMELS volumes are $25\hMpc$ across and contain few haloes above $10^{14}\,M_\odot$, so the densest regions we study are mostly galaxy- and group-scale haloes. We therefore present CosmoMap as a model for hot gas in haloes and the cosmic web; applying it to massive clusters requires training and testing on larger volumes (Section~\ref{sec:limitations}).

\section{Data and preprocessing}
\label{sec:data}

\subsection{CAMELS IllustrisTNG Latin hypercube}

We use the IllustrisTNG Latin-hypercube set of CAMELS \citep{VillaescusaNavarro2021CAMELS}: 1000 simulations, each following $256^3$ dark matter particles and $256^3$ initial gas cells in a periodic box of side $25\hMpc$, with a different value of
\begin{equation}
\mathbfit{p} = \{\Omega_{\mathrm{m}},\, \sigma_8,\, A_{\mathrm{SN1}},\, A_{\mathrm{AGN1}},\, A_{\mathrm{SN2}},\, A_{\mathrm{AGN2}}\},
\end{equation}
where $A_{\mathrm{SN1}}$ and $A_{\mathrm{SN2}}$ rescale the energy and wind speed of supernova feedback, and $A_{\mathrm{AGN1}}$ and $A_{\mathrm{AGN2}}$ the energy and burstiness of AGN feedback. The space-filling Latin-hypercube design matters here, because halo gas depends both on cosmology, through halo abundance and mass, and on feedback, through the gas thermodynamics. We chose TNG over SIMBA because its group and cluster gas has been compared extensively with X-ray observations and it includes a kinetic AGN feedback mode relevant to cluster cores. We use the 3D grids of the CAMELS Multifield Dataset \citep{VillaescusaNavarro2022Multifield} at $128^3$ resolution ($195\hkpc$ voxels) and $z \in \{0, 0.5, 1, 1.5, 2\}$. The inputs are the dark matter mass density $\rhodm$ and the magnitude of the dark matter velocity $|\mathbfit{v}_{\mathrm{DM}}|$; the targets are the gas temperature $T$ and the electron number density $\nel$. In these grids $T$ is the mass-weighted temperature of the gas in each voxel, and $\nel$ is the total number of electrons in the voxel divided by its comoving volume, in units of $h^2\,\mathrm{cm^{-3}}$ \citep{VillaescusaNavarro2022Multifield}; both are therefore averages over $195\hkpc$ voxels.

\subsection{Normalisation}

The density and gas fields span four to seven orders of magnitude and are close to log-normal, so we work with $x = \log_{10} X$, after applying small floors ($10$\,K for temperature and $10^{-12}\,h^2\,\mathrm{cm^{-3}}$ for $\nel$) that guard against zeros and lie well below the values in the data. Each field is then standardised with a single mean and standard deviation, computed once from a subset of the training simulations at $z=0$ and frozen:
\begin{equation}
x' = (x - \mu_X)/\sigma_X ,
\label{eq:norm}
\end{equation}
where $\mu_X$ and $\sigma_X$ are the mean and standard deviation of $x$ for field $X$. Because the same constants are used for every simulation and redshift, differences in amplitude between cosmologies and epochs are preserved, and predictions are converted to physical units with the same constants. No statistic of a target field is used at test time. The six parameters are standardised across the training set before conditioning.

\subsection{Data splitting}

We split the 1000 simulations once, with a fixed random seed, into 800 for training, 100 for validation and 100 for testing. All redshifts of a simulation belong to the same split, which prevents information leaking between splits through correlated snapshots. Every model and baseline uses this split, and field-level metrics are computed on the full $128^3$ volumes of the 100 test simulations; analyses that use subsets are noted where they appear.

\section{Model}
\label{sec:method}

\subsection{FiLM-conditioned U-Net}

CosmoMap is a 3D U-Net with three resolution levels of 64, 128 and 256 channels. Each level has a residual block of $3^3$ convolutions with group normalisation \citep[8 groups;][]{Wu2018GroupNorm} and spatial dropout ($p = 0.1$), and the bottleneck has two such convolutions. Encoder and decoder are joined by skip connections, and the network is fully convolutional. The same dark matter configuration produces different gas in different cosmologies and feedback models, so the network is conditioned on the parameters. The conditioning vector $\mathbfit{c}$ holds the six standardised parameters and the redshift. A two-layer perceptron maps it to a 192-dimensional embedding, from which every convolution in every block computes a channel-wise scale and shift,
\begin{equation}
\mathbfit{h} \rightarrow \left(1 + \gamma(\mathbfit{c})\right) \odot \mathbfit{h} + \beta(\mathbfit{c}),
\end{equation}
where $\mathbfit{h}$ is the feature map of the convolution, $\gamma$ and $\beta$ are the scale and shift, with one value per channel, and $\odot$ denotes channel-wise multiplication. The output head predicts a mean $\mu$ for each of $\log T$ and $\log\nel$ and, for the CosmoMap-$\sigma$ variant, a log-variance $\ln s^2$ for each, giving a Gaussian predictive distribution per voxel. Table~\ref{tab:model} summarises the configuration and the distribution of the $1.11 \times 10^7$ parameters. FiLM conditioning costs only 5 per cent of them. The network uses no attention layers: convolutions with a receptive field of several megaparsecs capture the local gravitational environment that sets the gas properties.

\begin{table*}
\centering
\caption{CosmoMap configuration and parameter distribution.}
\label{tab:model}
\small
\begin{tabular}{ll@{\hspace{2.5em}}lrr}
\toprule
\multicolumn{2}{c}{Configuration} & \multicolumn{3}{c}{Parameters by component} \\
\midrule
Input channels & 2 ($\log\rhodm$, $\log|\mathbfit{v}_{\mathrm{DM}}|$) & Decoder & 5\,671\,104 & 51.3\% \\
Outputs & $\mu$ (and $\ln s^2$) for $T$, $\nel$ & Bottleneck & 3\,540\,480 & 32.0\% \\
Levels / channels & 3 / 64, 128, 256 & Encoder & 1\,259\,200 & 11.4\% \\
Normalisation & GroupNorm, 8 groups & FiLM layers & 543\,488 & 4.9\% \\
Conditioning & 7 inputs $\rightarrow$ 192-d embedding & Conditioning MLP & 38\,592 & 0.3\% \\
Training patch & $64^3$ voxels ($12.5\hMpc$) & Input stem & 3\,648 & $<0.1$\% \\
Dropout & spatial, $p = 0.1$ & Output head & 260 & $<0.1$\% \\
\midrule
 & & Total & 11\,056\,772 & 100\% \\
\bottomrule
\end{tabular}
\end{table*}

\subsection{Loss functions}

The main CosmoMap model is trained with the mean squared error of the predicted means, averaged over voxels and both fields. This is the standard loss for a point estimate and weights all voxels equally. To obtain per-voxel uncertainties as well, the standard approach is to have the network also predict a variance and train it with a heteroscedastic Gaussian likelihood \citep{Kendall2017Uncertainties}. The CosmoMap-$\sigma$ variant does this with a composite loss that combines the heteroscedastic Gaussian negative log-likelihood with point-estimate terms and a variance penalty,
\begin{align}
\mathcal{L}_{\sigma} &= 0.408\,\mathcal{L}_{\mathrm{NLL}} + 0.150\,\mathcal{L}_{\mathrm{MSE}} + 0.134\,\mathcal{L}_{\mathrm{MAE}} \nonumber \\
&\quad + 0.207\,\mathcal{L}_{\mathrm{Huber}} + 0.01\,\langle s^2 \rangle, \\
\mathcal{L}_{\mathrm{NLL}} &= \tfrac{1}{2}\left\langle \frac{(y - \mu)^2}{s^2} + \ln s^2 \right\rangle,
\end{align}
where $y$ is the standardised target, $\mu$ and $s^2$ are the predicted mean and variance, angle brackets denote the average over voxels and both fields, and $\mathcal{L}_{\mathrm{MSE}}$, $\mathcal{L}_{\mathrm{MAE}}$ and $\mathcal{L}_{\mathrm{Huber}}$ are the mean squared, mean absolute and Huber errors of $\mu$. The weights are taken from a 67-trial search over the four weights, carried out in a different configuration with a different normalisation; they were not re-tuned here. Their precise values matter little here, because the ablation isolates the effect of the likelihood term directly (Section~\ref{sec:ablation}).

\subsection{Training and inference}

We train on random $64^3$ sub-volumes ($12.5\hMpc$), cut with periodic wrapping and augmented with the 48 symmetries of the cube, which are exact symmetries of the data because the velocity input is a magnitude. We use AdamW \citep{Loshchilov2019AdamW} with weight decay $10^{-5}$, batch size 4, gradient-norm clipping at 1 and mixed precision. The learning rate warms up linearly to $2\times 10^{-4}$ over the first epoch and then follows a cosine decay. Training runs for 30 epochs of 1000 steps, and we keep the checkpoint with the lowest loss on a fixed set of 320 validation sub-volumes (with one exception among the ablation variants, Section~\ref{sec:ablation}). CosmoMap and CosmoMap-$\sigma$ are close to convergence: in every seed the best checkpoint is at epoch 29 or 30, and the validation loss changes by less than 3 per cent over the last five epochs. The unconditioned and small networks select epochs 25 to 29, and their final validation losses are within 1 per cent of the best. One run takes about three hours on a single NVIDIA RTX 4070. CosmoMap, CosmoMap-$\sigma$, the unconditioned network and the small network (Section~\ref{sec:validation}) are each trained at $z=0$ with three random seeds; we report the mean and seed-to-seed standard deviation of their test metrics. Figures show the first seed, as do the analyses that need full predicted fields (scaling relations, structure detection and profiles, parameter recovery and parameter-space robustness). A second CosmoMap model is trained on all five redshifts.

To predict a full $128^3$ box we tile it with periodic $64^3$ windows, the size of the training sub-volumes, and keep the central $32^3$ of each, so that every predicted voxel has at least 16 voxels ($3.1\hMpc$) of context on each side. Matching the window to the training size keeps the group-normalisation statistics consistent with training. We adopt this scheme for that reason rather than for its test performance; it does not minimise the luminosity bias, the median $\log_{10}$ ratio of predicted to true box-integrated X-ray luminosity (Section~\ref{sec:xraymethod}; Appendix~\ref{app:inference}). Increasing the margin to 24 voxels changes no metric by more than $0.006$, so the predictions have converged with respect to context; Appendix~\ref{app:inference} compares other inference schemes. Once trained, the model predicts a full $128^3$ box in about 2\,s on the RTX 4070. Because the cosmological and feedback parameters enter only through the conditioning vector, one trained model predicts the gas for any parameter combination within the training range: a new cosmology or feedback model changes only this input and requires no retraining. For CosmoMap-$\sigma$, the aleatoric variance is the network output and the epistemic variance is the spread of the predicted means over 8 forward passes with dropout active \citep{Gal2016Dropout, Kendall2017Uncertainties}; their sum is the total predictive variance.

\section{Validation framework}
\label{sec:validation}

\subsection{Baselines and ablations}

We compare against three alternatives, all trained on the 800 training simulations:
\begin{description}
\item \textbf{Binned lookup:} the training-set mean of the target in bins of $\log\rhodm$ (width $0.1\sigma$), an estimate of the best one-to-one function of local dark matter density.
\item \textbf{Gradient-boosted trees:} a histogram gradient-boosted regressor \citep{Pedregosa2011Sklearn} for each target, on $\log\rhodm$ smoothed at 0, 1, 2, 4 and 8 voxels, $\log|\mathbfit{v}_{\mathrm{DM}}|$ at 0 and 2 voxels, the six parameters and the redshift. It is trained on $2.4\times10^6$ voxels drawn from all training simulations, with early stopping, and is converged with respect to both training data and boosting iterations (Appendix~\ref{app:baselines}). It sees the local multi-scale environment and the parameters, but has no learned spatial features.
\item \textbf{Unconditioned U-Net:} CosmoMap trained and evaluated with the conditioning vector set to zero, so that the FiLM layers reduce to fixed learned affine maps; it therefore has the same number of parameters.
\end{description}
The ablation study further retrains CosmoMap with a smaller network ($1.2\times10^6$ parameters, 20 base channels and 4 normalisation groups), with the point-estimate part of the composite loss only (MSE, MAE and Huber terms), and with three likelihood losses: the composite loss (CosmoMap-$\sigma$), the pure Gaussian negative log-likelihood $\mathcal{L}_{\mathrm{NLL}}$, and the $\beta$-NLL loss of \citet{Seitzer2022Pitfalls} with $\beta = 0.5$, which weights each voxel's likelihood term by $s^{2\beta}$ (with gradients stopped through the weight) so that poorly fitted voxels are down-weighted less.

\subsection{Field accuracy}

All metrics are computed on the 100 test boxes, which are used neither for training nor for model selection, and use $\log_{10}$ of the physical fields, so errors are fractional. For each test box we compute $R^2$, the root-mean-square error (RMSE) and the mean bias. To test the core objective, improved prediction in dense regions, we rank the voxels of each box by $\rhodm$ and compute the same metrics in six percentile ranges: deep voids ($<P_{10}$), voids ($P_{10}$--$P_{30}$), sheets ($P_{30}$--$P_{50}$), filaments ($P_{50}$--$P_{80}$), halo outskirts ($P_{80}$--$P_{95}$) and halo cores ($>P_{95}$). For scale dependence we compute the cross-correlation coefficient $r(k)$ and transfer function $\sqrt{P_{\mathrm{pred}}/P_{\mathrm{true}}}$ of the overdensity fields, where $k$ is the wavenumber, and $P_{\mathrm{pred}}$ and $P_{\mathrm{true}}$ are the power spectra of the prediction and the truth.

\subsection{Scaling relations}

Gravitational binding of gas in dark matter haloes gives power-law relations $\nel \propto \rhodm^{\alpha}$ and $T \propto \rhodm^{\beta}$. For each test box we fit $\alpha$ and $\beta$ by least squares in log space to $10^4$ random voxels, and separately to $10^4$ voxels from the densest five per cent, for the truth and for each model.

\subsection{Bright emission structures, profiles and luminosities}\label{sec:xraymethod}

We compute the bolometric free--free emissivity
\begin{equation}
\epsilon_X = A\,\nel^2\,T^{1/2}, \qquad A = 1.42 \times 10^{-27}\ \mathrm{erg\,s^{-1}\,cm^{3}\,K^{-1/2}},
\end{equation}
with a Gaunt factor of unity and the ion term $\sum_i Z_i^2 n_i$ replaced by $\nel$.\footnote{For fully ionised gas of primordial composition $\sum_i Z_i^2 n_i \approx 1.14\,\nel$, so this underestimates the emissivity by a constant factor that cancels in every comparison we make.} Because the fields are voxel averages, this is $\langle\nel\rangle^2\langle T\rangle^{1/2}$ rather than $\langle\nel^2 T^{1/2}\rangle$: it neglects sub-voxel clumping and is a resolution-dependent proxy for the X-ray emission of the simulated gas, not a prediction of its absolute luminosity. With $\nel$ in $h^2\,\mathrm{cm^{-3}}$ the emissivity also carries a factor $h^4$. Truth and prediction are treated identically, and we only compare them with each other, so the constant, the $h^4$ factor and the clumping factor cancel in all ratios. We identify bright structures as 26-connected regions above a fixed emissivity threshold, the 99.9th percentile of the true emissivity in 30 validation boxes. A true structure counts as recovered if a predicted structure overlaps it, and a predicted structure counts as real if it overlaps a true one; because any overlap is a lenient criterion that favours smoothed predictions, we also report a strict version that requires at least half of the voxels of the structure being tested (the true structure for completeness, the predicted one for purity) to lie inside bright structures of the other field. Around the three brightest emission peaks of each test box we extract surface brightness profiles by projecting the emissivity of a $41^3$-voxel cube along one axis, and fit the $\beta$-model
\begin{equation}
S_X(r) = S_0 \left[1 + (r/r_c)^2\right]^{-3\beta + 1/2}
\end{equation}
to the true and predicted profiles, where $r$ is the projected distance from the peak, $S_0$ the central surface brightness, $r_c$ the core radius and $\beta$ the slope parameter of the profile. We also compare the box-integrated emissivity $L$ and the emission-weighted temperature.

\subsection{Cosmological information}

To test whether cosmological information survives the mapping, we compute 32 summary statistics of each box's fields: the mean, standard deviation, skewness, kurtosis and 10th, 50th and 90th percentiles of $\log T$, $\log\nel$ and $\log\epsilon_X$; power spectrum amplitudes of $\nel$ and $\epsilon_X$ at $k = 1, 2, 5$ and $10\,h\,\mathrm{Mpc^{-1}}$; and the number and filling factor of bright emission regions and the volume fraction above $10^6$\,K. These statistics are intrinsic properties of the simulated 3D gas fields, not forward-modelled quantities such as X-ray fluxes: they test whether the predicted gas retains the parameter dependence of the simulated gas, not what an X-ray survey could constrain. A random forest with 500 trees regresses the six parameters, trained on the 100 validation boxes and tested on the 100 test boxes, so that no box used to train the networks is involved (the validation boxes were used only to select network checkpoints). With 100 training boxes for 32 features, the absolute $R^2$ values are limited by the size of the training set; the meaningful results are the comparisons between feature sources, which use identical forests. We compare four feature sources: the true gas fields (an upper bound for the predicted gas), the dark matter inputs alone (21 statistics: the same seven moments and percentiles of $\log\rhodm$ and $\log|\mathbfit{v}_{\mathrm{DM}}|$, power spectrum amplitudes of $\rhodm$ at the same four scales, and the number, filling factor and mean size of regions above a fixed density threshold), the unconditioned U-Net predictions, and the CosmoMap predictions. CosmoMap is given the true parameters, so recovery from its predictions is partly circular; the unconditioned network is the fair test of whether the mapping preserves information present in the dark matter.

\subsection{Uncertainty calibration and parameter-space robustness}

For CosmoMap-$\sigma$ we compare nominal and empirical coverage of central predictive intervals, then fit one multiplicative factor per field to the predicted standard deviations on the validation boxes and report test-set coverage before and after. To test robustness across parameter space, we compute the Mahalanobis distance $d_M = [(\mathbfit{p} - \boldsymbol{\mu})^{\mathsf{T}}\mathbfss{C}^{-1}(\mathbfit{p} - \boldsymbol{\mu})]^{1/2}$ of each test simulation from the training distribution, where $\boldsymbol{\mu}$ and $\mathbfss{C}$ are the mean and covariance matrix of the parameters of the training simulations, and correlate it with the test error. For draws from the prior, which is uniform in $\Omega_{\mathrm{m}}$, $\sigma_8$ and the logarithms of the feedback parameters, the mean distance is $d_M \approx 2.4$ and 98 per cent of draws lie between $1.2$ and $3.4$, so this tests uniformity across the prior volume rather than extrapolation beyond it. The distance is computed with $\Omega_{\mathrm{m}}$, $\sigma_8$ and the logarithms of the four feedback parameters, which span factors of 4 to 16.

\section{X-ray lightcone generation}
\label{sec:lightcone}

The intended application of CosmoMap is rapid generation of synthetic X-ray sky maps. We test this with the multi-redshift model, building $1^\circ \times 1^\circ$ lightcones on $256^2$ pixels from $z=0.05$ to $z = 2.25$ for 10 test simulations, each using its own value of $\Omega_{\mathrm{m}}$ in a flat $\Lambda$CDM background. The line of sight is cut into $25\hMpc$ slabs at comoving distance $\chi(z) = \int_0^z c\,\mathrm{d}z'/H(z')$. Each slab uses the snapshot nearest in redshift (the $z=2$ snapshot for $2 < z \le 2.25$), with a random cube symmetry and periodic shift; truth and prediction use identical transformations. Each slab's emissivity is projected along the line of sight and sampled at transverse comoving positions $\chi\theta$, with bilinear interpolation and periodic replication. Beyond $z \approx 0.55$ the field is wider than the box, so each slab is tiled two to three times across it and high-redshift structures recur within a map; this affects truth and prediction identically. The resulting surface brightness is
\begin{equation}
S = \frac{1}{4\pi (1+z)^4} \sum \epsilon_X\, \mathrm{d}l_{\mathrm{proper}},
\end{equation}
where the sum runs over the voxels of the slab along the line of sight, $z$ is the redshift of the slab, $\mathrm{d}l_{\mathrm{proper}}$ is the proper path length through one voxel, and the factor $(1+z)^{-4}$ is the cosmological dimming of surface brightness. We treat the gridded $\nel$ as a comoving density and convert it to physical density; this choice affects truth and prediction identically.

We then apply survey-like observational effects. The maps are convolved with a Gaussian point-spread function (PSF) of $30''$ full width at half maximum, comparable to the angular resolution of the eROSITA survey, and converted to counts so that the median pixel of the true map has two source counts. A uniform background of 40 counts per pixel is added, and Poisson noise is drawn using the same random numbers for truth and prediction, so that differences reflect only the predicted fields. Sources are detected as connected regions more than $5\sigma$ above a local, median-filtered background after smoothing with the PSF, and matched between maps within two PSF widths. Only relative fluxes are meaningful, because truth and prediction pass through the same pipeline and no instrument response, K-correction or absorption is applied. Once the boxes have been predicted, building one lightcone takes under a minute on a single CPU core.

\section{Results}
\label{sec:results}

\subsection{Field accuracy}

\begin{table*}
\centering
\caption{Test-set accuracy at $z=0$ (100 simulations, full $128^3$ volumes), in $\log_{10}$ physical units. Values are means over the 100 boxes; for the unconditioned U-Net, CosmoMap and CosmoMap-$\sigma$ they are further averaged over three training seeds, with the seed-to-seed standard deviation given. The last column is the box-to-box standard deviation of the mean bias of $\log T$, which measures how well each box's overall temperature level is predicted.}
\label{tab:main}
\small
\begin{tabular}{lcccccc}
\toprule
 & \multicolumn{2}{c}{$\log T$} & \multicolumn{2}{c}{$\log \nel$} & $\log \epsilon_X$ & $\sigma$(bias $T$) \\
Model & $R^2$ & RMSE [dex] & $R^2$ & RMSE [dex] & $R^2$ & [dex] \\
\midrule
Binned $\rhodm$ lookup & 0.173 & 0.671 & 0.652 & 0.307 & 0.631 & 0.210 \\
Gradient-boosted trees & 0.603 & 0.470 & 0.894 & 0.182 & 0.861 & 0.066 \\
U-Net, no conditioning & 0.647{\scriptsize$\,\pm0.008$} & 0.438{\scriptsize$\,\pm0.003$} & 0.906{\scriptsize$\,\pm0.001$} & 0.172{\scriptsize$\,\pm0.001$} & 0.877{\scriptsize$\,\pm0.0003$} & 0.142{\scriptsize$\,\pm0.002$} \\
CosmoMap-$\sigma$ & 0.733{\scriptsize$\,\pm0.004$} & 0.385{\scriptsize$\,\pm0.003$} & 0.922{\scriptsize$\,\pm0.001$} & 0.156{\scriptsize$\,\pm0.001$} & 0.904{\scriptsize$\,\pm0.002$} & 0.048{\scriptsize$\,\pm0.004$} \\
CosmoMap & $\mathbf{0.765}${\scriptsize$\,\pm0.001$} & $\mathbf{0.362}${\scriptsize$\,\pm0.001$} & $\mathbf{0.934}${\scriptsize$\,\pm0.0003$} & $\mathbf{0.142}${\scriptsize$\,\pm0.0003$} & $\mathbf{0.917}${\scriptsize$\,\pm0.0003$} & $\mathbf{0.030}${\scriptsize$\,\pm0.003$} \\
\bottomrule
\end{tabular}
\end{table*}

CosmoMap is the best model on every metric in Table~\ref{tab:main}. It reduces the RMSE of the gradient-boosted trees by 23 and 22 per cent for temperature and electron density, and that of the unconditioned network by 17 per cent for both fields, while its seed-to-seed variation of $R^2$ and RMSE is below 0.5 per cent. All models predict electron density much better than temperature: on these scales gas largely follows the dark matter potential, whereas temperature depends on the history of shocks, cooling and feedback heating.

The unconditioned U-Net improves on the trees, even though the trees see the parameters and the network does not, but most of CosmoMap's advantage comes from combining learned 3D spatial features with knowledge of the cosmology and feedback model. Conditioning also sets the overall level of each box: the box-to-box scatter of the mean temperature bias falls from $0.14\dex$ without conditioning to $0.03\dex$ with it. Since the normalisation is global, this level has to be inferred from the inputs and parameters.

\begin{figure*}
\centering
\includegraphics[width=\textwidth]{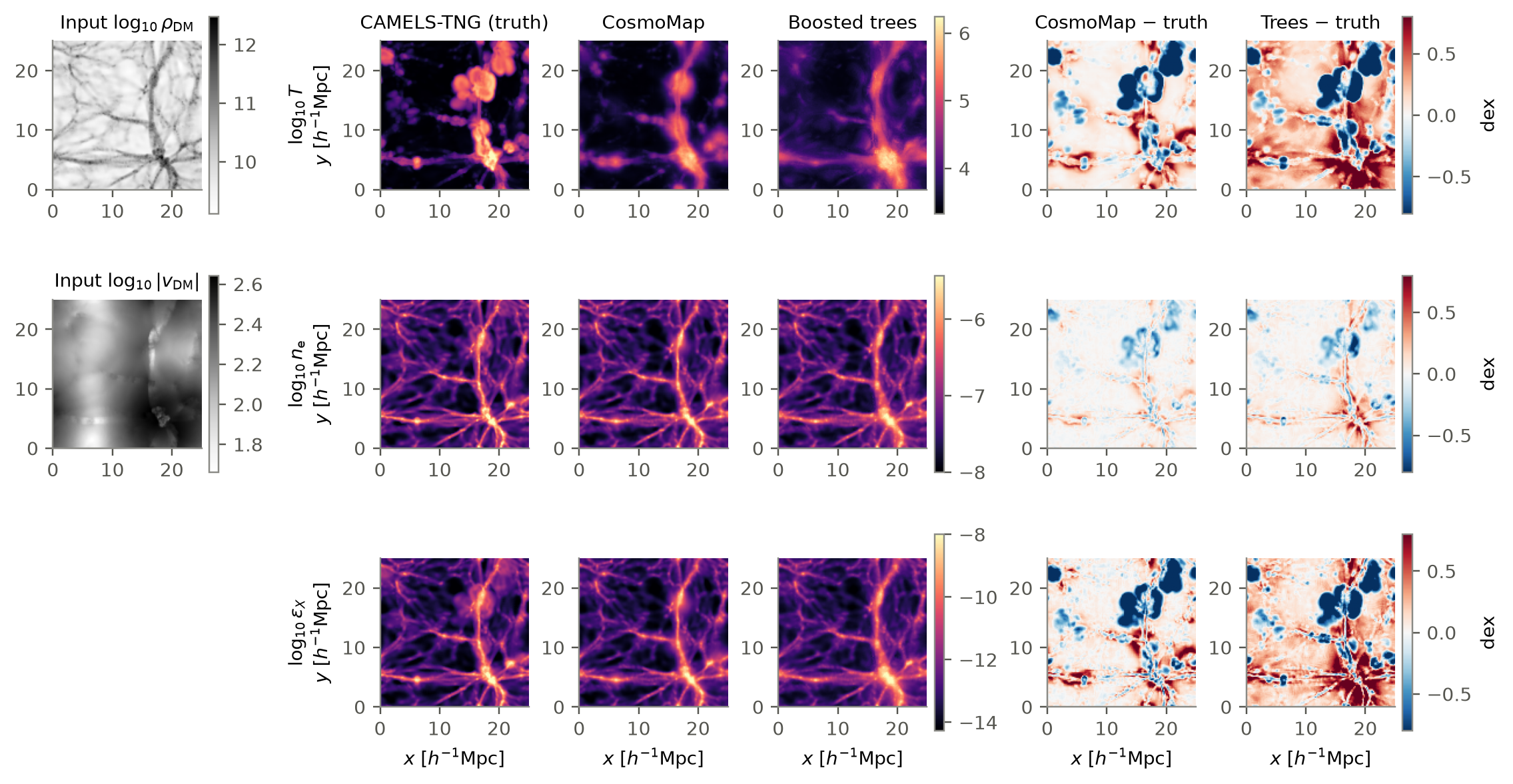}
\caption{A $1.56\hMpc$-thick slab (averaged in log space) through the test box with the median temperature $R^2$, at $z=0$. Rows show temperature, electron density and emissivity; columns show the inputs, the simulation, CosmoMap, the gradient-boosted trees, and the residuals of CosmoMap and of the trees.}
\label{fig:slices}
\end{figure*}

Figure~\ref{fig:slices} shows a slice through the test box with the median temperature $R^2$. Both CosmoMap and the trees recover the filaments and haloes, but CosmoMap reproduces the extent of the hot gas around haloes, heated by shocks and feedback, much better. It smooths the sharp outer edges of these regions, which is where its errors concentrate. The trees also systematically miss hot gas in regions that are cold on average, which CosmoMap largely places correctly.

\subsection{Ablation study}
\label{sec:ablation}

Table~\ref{tab:ablation} isolates the contribution of each component by retraining the $z=0$ model with one change at a time, under identical training and evaluation. Conditioning matters most: without FiLM, the temperature RMSE rises by 21 per cent and the box-to-box scatter of the luminosity rises from $0.24$ to $0.39\dex$ (seed averages). Model size matters less for the bulk of the volume: a network with nine times fewer parameters loses only $0.016$ in temperature $R^2$. It is clearly worse for the brightest gas, however: in every seed its electron density in the brightest one per cent of voxels (Section~\ref{sec:xray}) has $R^2 = 0.34$ against $0.49$, and its box luminosities are biased low by $0.41\dex$ on average, so capacity matters most where the X-ray emission is concentrated. Both reduced models also give box luminosities that depend on the training seed, from $-0.22$ to $-0.48\dex$ without conditioning and from $-0.24$ to $-0.51\dex$ for the small network, against $-0.15$ to $-0.17\dex$ for CosmoMap. Their voxel-level accuracy is reproducible to $0.01$ in $R^2$, but the level of the brightest gas is not.

The loss function has a large effect. Training with the composite likelihood loss (CosmoMap-$\sigma$) lowers the temperature $R^2$ from $0.765$ to $0.733$ and biases box-integrated luminosities low by $0.46\dex$, against $0.16\dex$ for the mean-squared-error loss. Both models are trained with three seeds, and both differences are seven to eight times their combined seed-to-seed standard deviation. Keeping the point-estimate terms but removing the likelihood term recovers most of the accuracy and removes most of the luminosity bias (one seed). Because the composite loss mixes the likelihood with other terms, with weights tuned in a different configuration, we also train with a pure Gaussian likelihood and with $\beta$-NLL (one seed each). The pure likelihood is the least accurate of the loss variants ($R^2 = 0.726$ and $0.917$) and gives the same luminosity bias as the composite loss ($-0.45\dex$). For $\beta$-NLL the validation loss is not a reliable guide to accuracy: in three training runs with the same seed, which differ because training on a GPU is not bit-reproducible, it is lowest at epoch 10, 13 or 15, while the validation RMSE of both fields falls by a further 2--5 per cent by the last epoch. For this variant we therefore evaluate the weights of the last epoch. $\beta$-NLL then recovers about 80 per cent of the temperature accuracy lost by the pure likelihood ($R^2 = 0.758$ in both runs that we evaluated at the last epoch) and removes part of the luminosity bias (leaving $-0.33$ and $-0.25\dex$). The checkpoints with the lowest validation loss are less accurate ($R^2 = 0.740$--$0.749$), and their luminosity bias ranges from $-0.22$ to $-0.50\dex$, so the luminosity bias of this variant is not well determined. The cost of learning per-voxel variances is therefore not an artefact of the composite loss, although $\beta$-NLL reduces it. This is consistent with a known failure mode of heteroscedastic regression \citep{Seitzer2022Pitfalls}: the likelihood lets the network assign large variance to voxels it fits poorly, which down-weights them in training. The accuracy cost is spread over all environments, with temperature $R^2$ lower by $0.05$ in deep voids and $0.03$ in the densest regime. Because the X-ray luminosity is dominated by the densest gas, however, the under-prediction there likely produces the large luminosity bias. The likelihood-trained model remains useful for its per-voxel uncertainties (Section~\ref{sec:calib}).

\begin{table*}
\centering
\caption{Ablation study at $z=0$: test-set accuracy and median log ratio of predicted to true box luminosity when one component is changed. CosmoMap, CosmoMap-$\sigma$ and the networks without conditioning or with fewer parameters are averaged over three seeds, with the seed-to-seed standard deviation of the luminosity ratio given; the other variants use one seed. The $\beta$-NLL row uses the weights of the last epoch (Section~\ref{sec:ablation}).}
\label{tab:ablation}
\small
\begin{tabular}{lrccccc}
\toprule
 & & \multicolumn{2}{c}{$\log T$} & \multicolumn{2}{c}{$\log\nel$} & \\
Configuration & Params & $R^2$ & RMSE & $R^2$ & RMSE & $\log\frac{L_{\mathrm{pred}}}{L_{\mathrm{true}}}$ \\
\midrule
CosmoMap (MSE loss) & 11.1\,M & 0.765 & 0.362 & 0.934 & 0.142 & $-0.16${\scriptsize$\,\pm0.01$} \\
No FiLM conditioning & 11.1\,M & 0.647 & 0.438 & 0.906 & 0.172 & $-0.37${\scriptsize$\,\pm0.14$} \\
Small network & 1.2\,M & 0.749 & 0.373 & 0.924 & 0.154 & $-0.41${\scriptsize$\,\pm0.15$} \\
MSE + MAE + Huber loss & 11.1\,M & 0.759 & 0.366 & 0.935 & 0.142 & $-0.19$ \\
Composite likelihood loss (CosmoMap-$\sigma$) & 11.1\,M & 0.733 & 0.385 & 0.922 & 0.156 & $-0.46${\scriptsize$\,\pm0.04$} \\
Pure Gaussian likelihood & 11.1\,M & 0.726 & 0.390 & 0.917 & 0.161 & $-0.45$ \\
$\beta$-NLL loss ($\beta = 0.5$) & 11.1\,M & 0.758 & 0.367 & 0.931 & 0.146 & $-0.33$ \\
\bottomrule
\end{tabular}
\end{table*}

\subsection{Performance in high-density regions}

\begin{figure*}
\centering
\includegraphics[width=\textwidth]{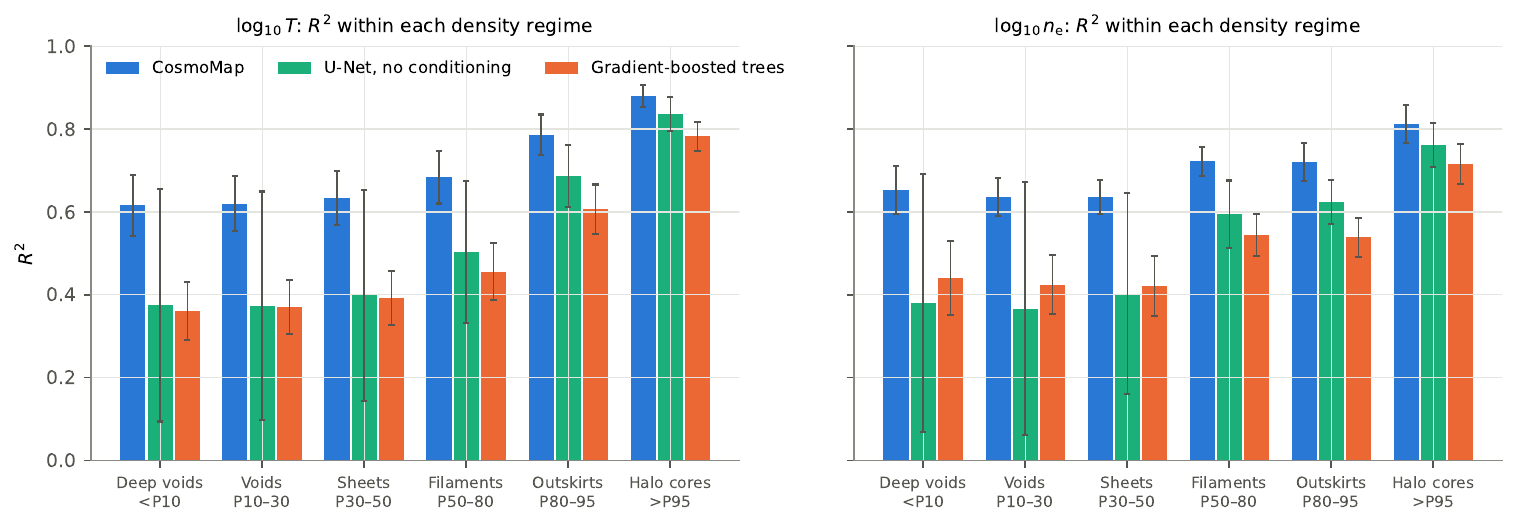}
\caption{$R^2$ of temperature (left) and electron density (right) within six ranges of dark matter density at $z=0$. Error bars show the scatter over the 100 test boxes. The binned lookup, not shown, has negative $R^2$ in every regime except for electron density in cores ($0.35$).}
\label{fig:regimes}
\end{figure*}

The core objective of this work is accurate prediction in the densest environments. We label the densest five per cent of voxels ``halo cores''; in $25\hMpc$ boxes these are mostly the central regions of galaxy- and group-scale haloes, with few true clusters. Figure~\ref{fig:regimes} shows that temperature accuracy improves systematically with density: for CosmoMap (averaged over seeds), $R^2$ rises from $0.62$ in deep voids to $0.88$ in halo cores, and the RMSE falls from $0.38$ to $0.24\dex$, so the gain is not merely an effect of the larger dynamic range in dense regions. In cores, gas is expected to be heated mainly by gravitational collapse and accretion shocks, which the dark matter traces. At low densities, temperature depends on photoheating, adiabatic cooling and the reach of feedback-driven outflows, which local dark matter constrains less.

Conditioning shows the same physics from the other side. In cores the learned models are within about $0.1$ in $R^2$ of each other ($0.88$, $0.84$ and $0.78$ for CosmoMap, the unconditioned network and the trees), while in voids and sheets CosmoMap reaches $R^2 \approx 0.63$ against $0.39$ for the unconditioned network and $0.38$ for the trees. A natural interpretation is that this is where the gas temperature depends most on how far and how strongly feedback has heated it, and only the conditioned model knows the feedback parameters. Electron density follows a similar pattern.

CosmoMap's mean biases are at most $0.02\dex$ in every regime, becoming slowly more negative towards the densest regions, while the unconditioned network over-predicts electron density by $0.01$--$0.03\dex$, most in voids, likely reflecting the ambiguity in the feedback strength of a given simulation (Fig.~\ref{fig:bias} in Appendix~\ref{app:figs}).

In Fourier space (Fig.~\ref{fig:power} in Appendix~\ref{app:figs}), CosmoMap's electron density and emissivity cross-correlate with the truth at $r(k) > 0.9$ down to scales of about $1\hMpc$ ($k \approx 7$--$8\,h\,\mathrm{Mpc^{-1}}$), and the density amplitude is recovered to within 7 per cent up to $k \approx 5\,h\,\mathrm{Mpc^{-1}}$; the trees decorrelate beyond $k \approx 2$--$3\,h\,\mathrm{Mpc^{-1}}$. Temperature decorrelates on larger scales ($r > 0.9$ only to $k \approx 1\,h\,\mathrm{Mpc^{-1}}$), with excess power at $k \approx 1$--$2\,h\,\mathrm{Mpc^{-1}}$ and a deficit on small scales, consistent with the smoothed edges of the hot regions.

\subsection{Redshift evolution}

Table~\ref{tab:redshift} compares the multi-redshift model with baselines trained on the same five redshifts. To limit the cost of computing features for all snapshots, these baselines use 160 of the training simulations, which at $z=0$ lowers the tree $R^2$ by about $0.02$ for temperature and $0.006$ for density relative to Table~\ref{tab:main}. All models predict both fields with smaller RMSE at higher redshift, where structures are less evolved and feedback has had less time to act. CosmoMap is the best model at every redshift: its temperature RMSE is 24--29 per cent below that of the trees, with the largest gain at $z=2$, and its density RMSE 22--23 per cent below. At $z=0$ the multi-redshift model is almost as accurate as the $z=0$ model ($R^2 = 0.760$ against $0.765$ for temperature), even though the same training budget is spread over five redshifts.

\begin{table*}
\centering
\caption{Test-set accuracy of the multi-redshift CosmoMap model and the baselines trained on the same redshifts.}
\label{tab:redshift}
\small
\begin{tabular}{lcccccc}
\toprule
 & \multicolumn{3}{c}{$\log T$: $R^2$ (RMSE [dex])} & \multicolumn{3}{c}{$\log \nel$: $R^2$ (RMSE [dex])} \\
$z$ & Binned & Trees & CosmoMap & Binned & Trees & CosmoMap \\
\midrule
0.0 & 0.158 (0.674) & 0.581 (0.482) & \textbf{0.760} (0.366) & 0.645 (0.306) & 0.888 (0.187) & \textbf{0.931} (0.146) \\
0.5 & 0.249 (0.558) & 0.594 (0.412) & \textbf{0.777} (0.306) & 0.657 (0.269) & 0.915 (0.149) & \textbf{0.949} (0.115) \\
1.0 & 0.309 (0.449) & 0.617 (0.336) & \textbf{0.794} (0.247) & 0.659 (0.240) & 0.938 (0.117) & \textbf{0.963} (0.090) \\
1.5 & 0.341 (0.365) & 0.650 (0.271) & \textbf{0.815} (0.197) & 0.656 (0.221) & 0.954 (0.094) & \textbf{0.972} (0.073) \\
2.0 & 0.364 (0.285) & 0.700 (0.204) & \textbf{0.849} (0.144) & 0.647 (0.203) & 0.967 (0.073) & \textbf{0.980} (0.057) \\
\bottomrule
\end{tabular}
\end{table*}

\subsection{Scaling relations}

The predicted fields reproduce the dark matter--gas scaling relations. Over all voxels, the true slopes are $\alpha = 0.955 \pm 0.029$ and $\beta = 0.642 \pm 0.075$ (mean and box-to-box scatter), and CosmoMap and the trees give $0.948$ and $0.631$, and $0.952$ and $0.640$; slopes fitted to random voxels are dominated by voids and sheets, so this test barely discriminates between models. In the densest five per cent of voxels the slopes flatten to $\alpha = 0.796$ and $\beta = 0.552$; CosmoMap gives $0.783$ and $0.547$, with half the per-box error in $\alpha$ of the trees ($0.019$ against $0.038$), and tracks the variation of the slopes between boxes. These are voxel-level relations that mix gas in haloes, filaments and voids, not the halo scaling relations used in cluster cosmology, and their dependence on the feedback parameters need not follow physical expectations for haloes; we use them only to compare predicted and true fields from the same simulations, and their agreement does not imply that individual haloes are recovered accurately (Section~\ref{sec:xray}).

\subsection{X-ray emissivity, bright structures and profiles}\label{sec:xray}

Emissivity is predicted well voxel by voxel ($R^2 = 0.917$ in log space), but voxel-level scores are dominated by the far more numerous low-density voxels and say little about clusters, so we test the bright structures directly. Projected emissivity maps correlate with the truth at $r = 0.976$ (box-to-box standard deviation $0.007$). Of the 17\,416 bright structures in the true test boxes, CosmoMap recovers 76 per cent, the unconditioned network 76 per cent and the trees 23 per cent, and 84, 80 and 68 per cent of their predicted structures are real (pooled over boxes). With the strict criterion, completeness falls to 66, 65 and 13 per cent and purity to 82, 74 and 59 per cent, so CosmoMap is as complete as the unconditioned network, far more complete than the trees, and its structures are more often real than those of either alternative. The luminosity of the recovered structures is low by a median of $0.21\dex$ for CosmoMap, $0.20\dex$ for the unconditioned network and $0.76\dex$ for the trees. The reason is visible in the brightest one per cent of voxels: there temperature is accurate ($R^2 = 0.84$, bias $-0.04\dex$), but electron density is biased low by $0.12\dex$ ($R^2 = 0.49$). The emission-weighted temperature of each box, a proxy for the spectroscopic temperature, has a median offset within $0.03\dex$ for every seed, compared with $0.52\dex$ too low for the trees.

\begin{figure}
\centering
\includegraphics[width=\columnwidth]{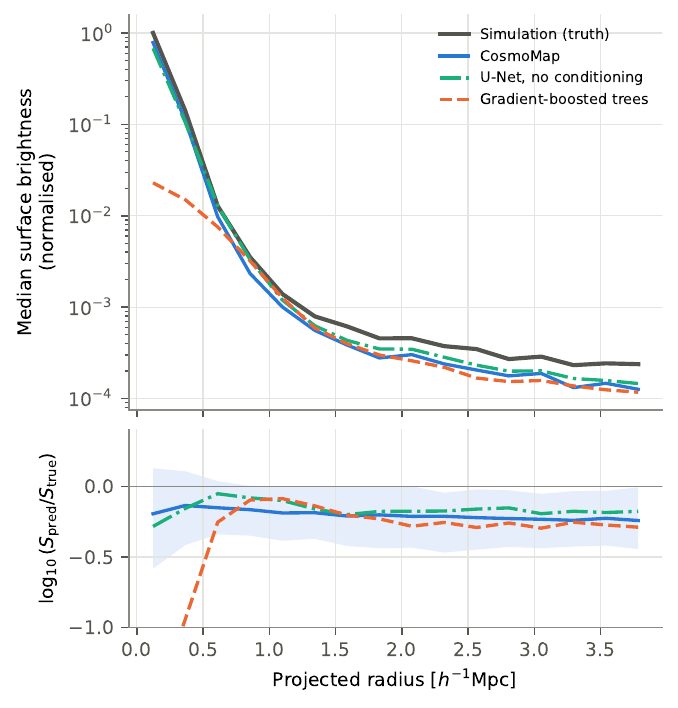}
\caption{Median surface brightness profiles around the three brightest emission peaks of each test box (295 profiles, $z=0$), normalised to the central truth value (top), and the median log ratio of predicted to true profile (bottom; the shaded band is the 16--84 per cent range for CosmoMap).}
\label{fig:sb}
\end{figure}

\begin{figure*}
\centering
\includegraphics[width=\textwidth]{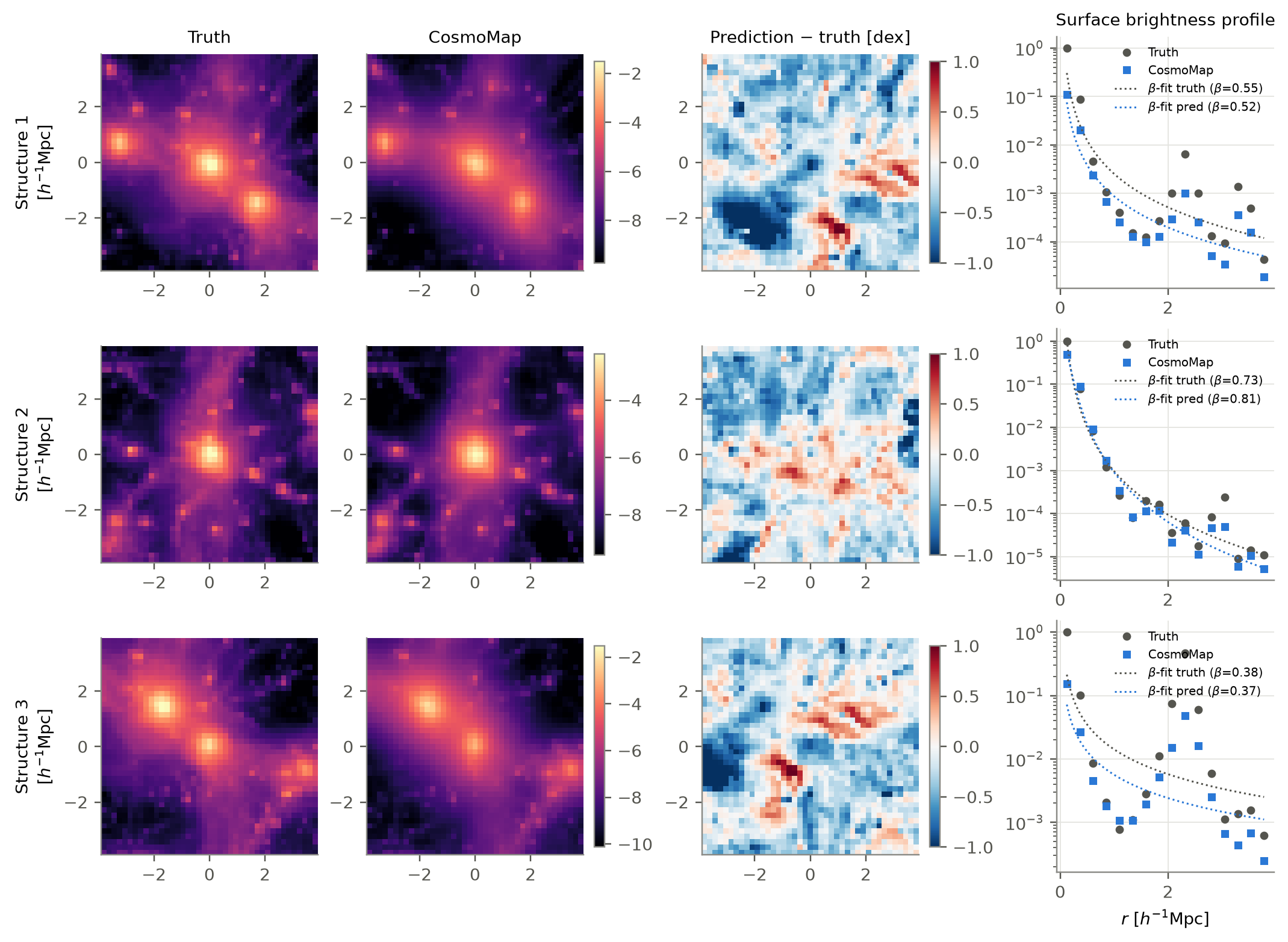}
\caption{Projected emissivity around the three brightest structures in one test box at $z=0$: truth, CosmoMap and residual (left), and surface brightness profiles with $\beta$-model fits (right). The maps are $8.0\hMpc$ across and integrated over the same depth.}
\label{fig:clusters}
\end{figure*}

Figure~\ref{fig:sb} compares surface brightness profiles around the three brightest peaks of each test box (295 profiles, because some boxes have fewer than three peaks above the emissivity threshold). The predicted profiles have the right shape: fitted $\beta$-model slopes have medians of $0.603$ (truth) and $0.611$ (CosmoMap) and correlate at $r = 0.80$ from structure to structure. The predicted profiles lie below the truth at all radii. The deficit is $0.19\dex$ in the central bin and $0.13\dex$ at $0.4\hMpc$, and then grows outwards to $0.24\dex$ at $3$--$4\hMpc$; beyond $1\hMpc$ 84--89 per cent of the profiles lie below the truth. Outside the central bin, the outskirts are therefore under-predicted more than the inner regions, consistent with the under-predicted extent of the hot gas around haloes (Fig.~\ref{fig:slices}). The core radius is not measurable at this resolution: for 88 per cent of the true profiles the fit prefers $r_c$ smaller than one voxel, so the fitted $\beta$ effectively measures the slope of the outer profile. Figure~\ref{fig:clusters} shows three individual structures. The emission morphology, including substructure and neighbouring haloes, is reproduced, and the residual maps show the main failure mode: hot regions whose extent is under-predicted.

\begin{figure}
\centering
\includegraphics[width=\columnwidth]{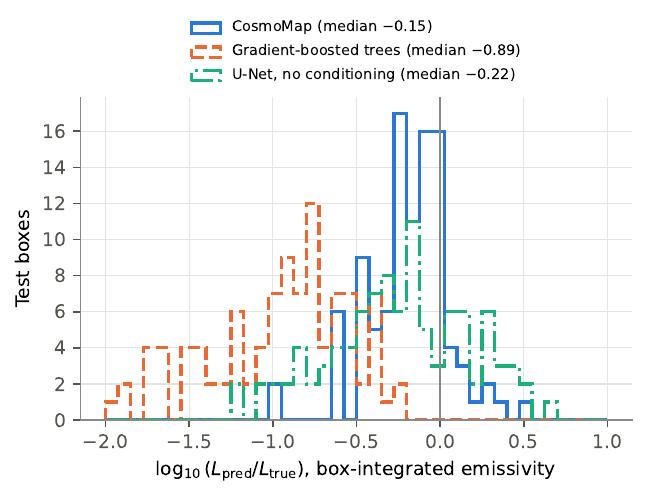}
\caption{Ratio of predicted to true box-integrated emissivity for the 100 test boxes at $z=0$ (first seed of each network).}
\label{fig:lum}
\end{figure}

Box-integrated luminosities have a median $\log_{10}(L_{\mathrm{pred}}/L_{\mathrm{true}})$ of $-0.16$ (seed average; $-0.15$, $-0.17$ and $-0.17$ for the three seeds), with a box-to-box scatter of $0.24\dex$ (Fig.~\ref{fig:lum}). For comparison, the trees give $-0.89\dex$ with $0.42\dex$ scatter, the unconditioned network $-0.37\dex$ (seed average; $-0.22$ to $-0.48\dex$) with $0.39\dex$, and CosmoMap-$\sigma$ $-0.46\dex$ with $0.33\dex$. For CosmoMap-$\sigma$, part of the bias can be removed after training. If each voxel's predictive distribution of $\log_{10}\epsilon_X$ is Gaussian with variance $s^2 = 4 s_{n_\mathrm{e}}^2 + s_T^2/4$, the expected emissivity is $10^{\mu}\exp[\tfrac{1}{2}(\ln 10)^2 s^2]$; for the first seed this correction brings the median to $-0.04\dex$ with the raw variances and to $-0.20\dex$ with the validation-calibrated variances (Section~\ref{sec:calib}), without reducing the scatter. These luminosity biases are not robust at the $0.2\dex$ level: across inference schemes the median bias of CosmoMap ranges from $-0.15$ to $+0.04\dex$ (first seed), while the voxel-level metrics barely change (Section~\ref{sec:limitations} and Appendix~\ref{app:inference}).

\subsection{Cosmological parameter recovery}

\begin{figure*}
\centering
\includegraphics[width=\textwidth]{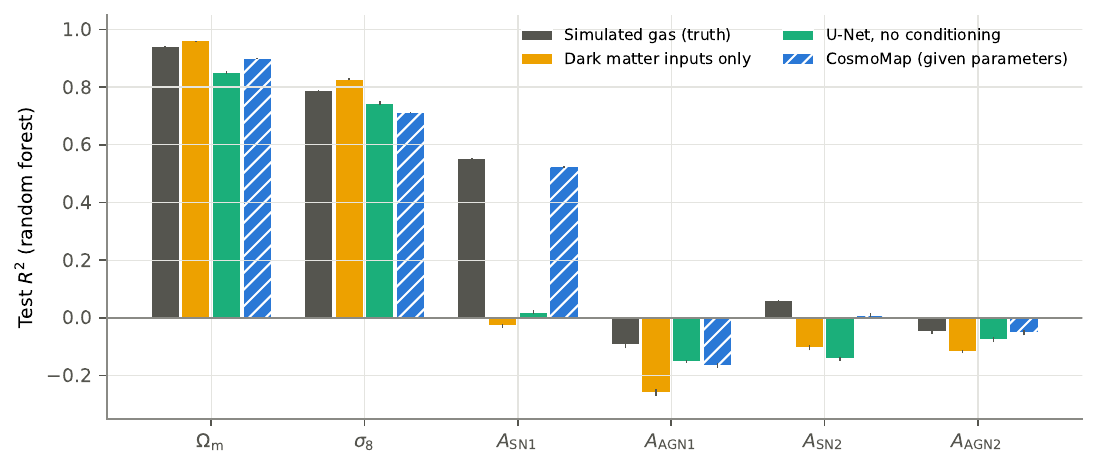}
\caption{Test $R^2$ of a random forest predicting the six CAMELS parameters from summary statistics of each field source, trained on validation boxes and tested on test boxes at $z=0$. Error bars show the scatter over five forest seeds. Recovery from CosmoMap predictions (hatched) is partly circular, because the network is conditioned on the true parameters.}
\label{fig:paramrec}
\end{figure*}

Figure~\ref{fig:paramrec} shows how well the six parameters can be recovered from summary statistics. From the true gas fields, $\Omega_{\mathrm{m}}$ and $\sigma_8$ are recovered with $R^2 = 0.94$ and $0.79$, and of the feedback parameters only the supernova energy $A_{\mathrm{SN1}}$ ($R^2 = 0.55$). The dark matter inputs alone give $0.96$ and $0.83$ for the cosmological parameters, showing that most of this information is already present in the dark matter. Predictions of the unconditioned network preserve most of it ($R^2 = 0.85$ and $0.74$) but, as expected, contain no information on feedback. Predictions of CosmoMap give $0.90$ and $0.71$ for the cosmological parameters and $0.52$ for $A_{\mathrm{SN1}}$, close to the true gas for $A_{\mathrm{SN1}}$. As a ratio, $\Omega_{\mathrm{m}}$ recovered from the CosmoMap predictions is $\Omega_{\mathrm{m,pred}}/\Omega_{\mathrm{m,true}} = 1.01 \pm 0.11$ (median and half the 16--84 per cent range over the test boxes), against $1.01 \pm 0.09$ from the true gas and $1.01 \pm 0.15$ from the unconditioned network; for $\sigma_8$ the values are $1.02 \pm 0.08$, $1.01 \pm 0.07$ and $1.01 \pm 0.08$. Because CosmoMap is given the parameters, its recovery of $A_{\mathrm{SN1}}$ shows that the network imprints feedback on the gas with a realistic signature, not that the information was recovered from the dark matter. None of the remaining feedback parameters is recoverable from either true or predicted fields at this resolution.

The median relative error of the recovered $\Omega_{\mathrm{m}}$ and $\sigma_8$ is 4--11 per cent whichever fields the summary statistics are computed from (true gas, dark matter, unconditioned network or CosmoMap; Fig.~\ref{fig:paramscatter} in Appendix~\ref{app:figs} shows three of these), and their mean biases are at most 3 per cent of the parameter range. These constraints are still several times looser than the few-per-cent precision targeted by cluster cosmology surveys. The other feedback parameters reach training $R^2$ of $0.6$--$0.8$ but test $R^2$ near zero, so their apparent signal is over-fitting.

\subsection{Uncertainty quantification}
\label{sec:calib}

Per-voxel uncertainties come from CosmoMap-$\sigma$ and are dominated by the aleatoric part: Monte Carlo dropout contributes only 3 and 4 per cent of the variance for temperature and density (first seed), although it is known to underestimate epistemic uncertainty.

\begin{figure}
\centering
\includegraphics[width=\columnwidth]{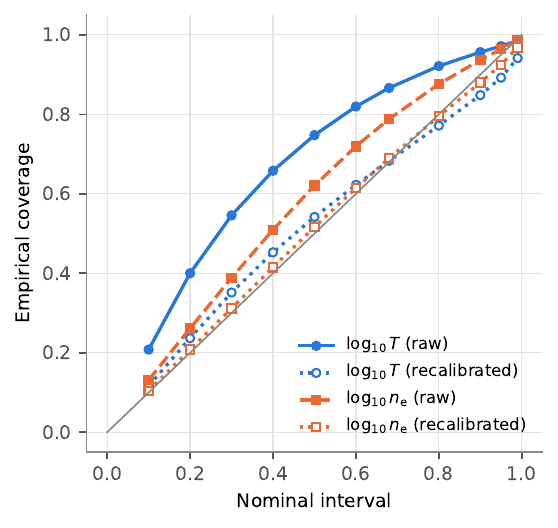}
\caption{Empirical coverage of central predictive intervals of CosmoMap-$\sigma$ against nominal level for the 100 test boxes at $z=0$ (every 64th voxel, first seed), with the raw uncertainties and after rescaling by one factor per field fitted on the 100 validation boxes.}
\label{fig:calib}
\end{figure}

We assess the uncertainties through their coverage: the fraction of test voxels whose true value lies inside the central predictive interval of a given nominal probability, for example the $\pm1\sigma$ interval for 68 per cent. Well-calibrated uncertainties give a coverage equal to the nominal probability, so the curves in Fig.~\ref{fig:calib} would lie on the diagonal. The raw 68 per cent intervals are too wide, containing 84 and 78 per cent of temperature and density voxels (seed average), while the raw 95 per cent intervals contain 97 and 96 per cent. One scale factor per field, fitted on the validation boxes, brings the 68 per cent coverage to $0.67$--$0.69$, but the 95 per cent coverage then falls to $0.89$ and $0.92$, because the errors have heavier tails than a Gaussian. Given the accuracy cost of the likelihood loss, a practical approach is to use CosmoMap for the fields and CosmoMap-$\sigma$ for their uncertainties. Scored against the residuals of CosmoMap trained with the same seed, the CosmoMap-$\sigma$ uncertainties rank the errors as well as they rank their own (Spearman correlation with the absolute residual $0.77$--$0.79$ for temperature and $0.62$ for density) and, with the scale factor refitted, reach the same coverage.

These coverages are dominated by voids and sheets. In the brightest one per cent of emission voxels the globally rescaled 68 per cent intervals contain only 55 and 57 per cent of temperature and density voxels (first seed), and the predicted standard deviation barely ranks the errors (Spearman correlation $0.17$--$0.18$). Rescaling instead with one factor per field in five bins of predicted emissivity, which needs no truth at test time, gives a 68 per cent coverage of $0.69$ over all voxels and, in the brightest one per cent, $0.70$ for temperature and $0.62$ for electron density. The density uncertainties of the brightest gas therefore remain somewhat too narrow, and such an environment-dependent rescaling should be applied before the uncertainties are propagated into mock X-ray maps.

\subsection{Robustness across parameter space}

The test simulations lie at Mahalanobis distances between $1.4$ and $3.5$ from the centre of the training distribution, and the test error is uncorrelated with this distance ($r = 0.02$ and $0.01$ for temperature and density), so performance does not degrade towards the edges of the prior volume. It does depend on the individual parameters (Fig.~\ref{fig:params} in Appendix~\ref{app:figs}): errors grow with $\Omega_{\mathrm{m}}$ and $\sigma_8$ (correlation coefficients $0.43$--$0.60$), which produce more collapsed, shock-heated structure, and fall with the supernova energy $A_{\mathrm{SN1}}$ ($-0.58$ for temperature), whose stronger winds smooth the gas. Because the errors are in $\log_{10}$, it is the fractional error that grows with $\Omega_{\mathrm{m}}$ and $\sigma_8$. The other feedback parameters have weaker effects ($|r| \le 0.30$).

\subsection{Gravity-only inputs}
\label{sec:nbody}

In an application, the input dark matter would come from a gravity-only simulation rather than from the hydrodynamical run, whose dark matter responds to the baryons. To test this, we replace both inputs of each test box by those of the matched CAMELS $N$-body run, which shares its initial conditions and parameters. The density is converted from mass per voxel to density and rescaled by $(\Omega_{\mathrm{m}}-\Omega_{\mathrm{b}})/\Omega_{\mathrm{m}}$ (with $\Omega_{\mathrm{b}} = 0.049$) to the dark matter share. The gridded $N$-body data provide only the density, so we grid the velocities from the $z=0$ $N$-body snapshots: each particle's speed is deposited with its mass using the fourth-order (piecewise cubic spline) kernel of the Multifield Dataset, and the mass-weighted mean speed is taken in each voxel. Applied to the dark matter of a hydrodynamical run, this procedure reproduces the published velocity grid of a test box (log correlation $0.998$ over all voxels, median deviation $0.7$ per cent). Without retraining, the temperature and electron density $R^2$ change from $0.764$ and $0.934$ to $0.760$ and $0.930$ (first seed), and the median box luminosity from $-0.15$ to $-0.22\dex$. Replacing only the density gives nearly the same result ($0.758$, $0.930$ and $-0.24\dex$), so the velocity input is barely affected by the baryons. Baryonic effects on the dark matter therefore change the voxel-level predictions very little, although they have a measurable effect on the brightest structures.

\subsection{Mock X-ray lightcones}

\begin{figure*}
\centering
\includegraphics[width=\textwidth]{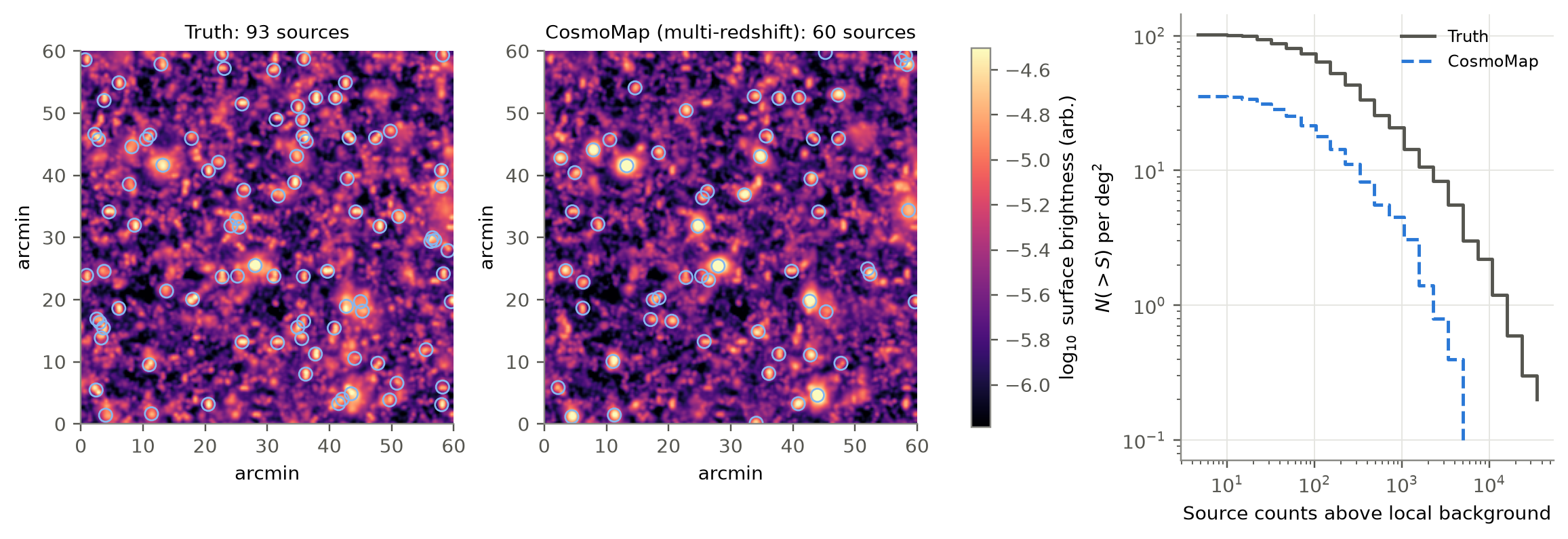}
\caption{A $1^\circ\times1^\circ$ mock X-ray lightcone ($0.05 < z < 2.25$) built from the true fields (left) and from the multi-redshift CosmoMap predictions (centre), after convolution with a $30''$ PSF; circles mark detected sources. This is the first of the 10 lightcones; it has 93 true and 60 predicted sources, against 1016 and 357 over all ten. Right: cumulative source counts above the local background, per square degree, over 10 lightcones.}
\label{fig:lc}
\end{figure*}

Figure~\ref{fig:lc} compares a lightcone built from the true fields with one built from CosmoMap's predictions. Averaged over 10 lightcones, the PSF-convolved maps correlate at $r = 0.95$ in log surface brightness, the large-scale pattern of emission is reproduced, and the total flux is low by $0.30\dex$. Source detection is harder: at the same detection threshold, the predicted maps contain 357 sources against 1016 in the true maps. Pooled over the lightcones, 28 per cent of the true sources are recovered (281 of 1016), and 80 per cent of the predicted sources (287 of 357) correspond to true ones, although this purity ranges from 40 to 95 per cent between lightcones; one in five predicted sources is therefore a false positive. Completeness rises steeply with flux: CosmoMap recovers 46 per cent of the true sources brighter than the median, 62 per cent of the brightest quarter and 89 per cent of the brightest tenth (102 sources, with fluxes above about 11 times the median), and 94 per cent of the brightest tenth of its own detections are real. Matched sources are fainter by a median of $0.84\dex$. The model misses mainly the fainter emission peaks that are detectable in the true maps, consistent with its smoothing of compact structures. In its current form CosmoMap is therefore well suited to the diffuse and large-scale X-ray signal, but source counts require a model that reproduces compact peaks more faithfully. The detected sources here are emission peaks from simulated groups and clusters, not a realistic X-ray source population.

\section{Discussion}
\label{sec:discussion}

\subsection{Accuracy versus uncertainty}

The ablation exposes a trade-off: a heteroscedastic likelihood gives useful per-voxel uncertainties but, by down-weighting poorly fitted voxels, under-fits the densest gas and biases luminosities, and $\beta$-NLL \citep{Seitzer2022Pitfalls}, designed to reduce this down-weighting, recovers most of the accuracy and removes part of the luminosity bias. All three likelihood variants over-cover before rescaling (raw 68 per cent coverage $0.82$--$0.87$ for temperature and $0.77$--$0.81$ for electron density, first seed) and reach $0.68$--$0.70$ after one rescaling. Training the variance head after freezing a mean-trained network could provide uncertainties without the accuracy cost; the pairing test of Section~\ref{sec:calib} suggests that this would work.

\subsection{Limitations}
\label{sec:limitations}

\paragraph*{Compact structures.} CosmoMap smooths the sharpest structures, so the electron density of the brightest gas is under-predicted by about $0.12\dex$ and nearly three quarters of the sources detected in the true lightcones are missed. This smoothing is expected from the squared-error loss, which is minimised by the average of all gas configurations consistent with the dark matter and so blurs peaks whose position and amplitude the input does not fix. Reproducing them probably requires a model that draws samples of the gas field instead of predicting its mean, such as a diffusion model or a generative adversarial network.

\paragraph*{Inputs from the hydrodynamical run.} The network is trained on dark matter fields from the same simulations as the targets, which include the back-reaction of baryons on the dark matter. With the density and velocity of matched gravity-only runs as input, the accuracy drops only slightly (Section~\ref{sec:nbody}), but box luminosities fall by a further $0.07\dex$. Training on gravity-only inputs would remove this mismatch.

\paragraph*{Normalisation statistics.} Group normalisation computes its statistics over the input volume, so the predictions depend on the size of the inference window, which is why we use windows of the training size. With full $128^3$ boxes instead, the voxel-level $R^2$ of CosmoMap changes by at most $0.005$, but its median box luminosity shifts from $-0.15$ to $+0.03\dex$ (first seed), and the unconditioned network, which must infer the overall gas level from the input statistics, loses $0.05$ in density $R^2$. Normalisation with statistics fixed after training removes this sensitivity. Retrained with batch normalisation \citep{Ioffe2015BatchNorm} and otherwise unchanged (first seed), CosmoMap reaches $R^2 = 0.761$ and $0.935$ with a median box luminosity of $-0.18\dex$ for every inference scheme we tested, including the full box (Table~\ref{tab:inference}), and the box-to-box scatter of the luminosity falls from $0.25$ to $0.17\dex$. Its accuracy matches that of the group-normalised model, so it can replace it where the inference window must vary; we keep the group-normalised model as the reference because all other experiments use it.

\paragraph*{Resolution and volume.} At $195\hkpc$ per voxel, the cores of haloes are not resolved (the fitted core radii are below one voxel), and features such as cool cores and AGN cavities are absent. A $25\hMpc$ box contains few haloes above $10^{14}\,M_\odot$, so the halo-core regime is dominated by galaxy- and group-scale haloes, and the results should be extended to cluster masses with larger volumes before being applied to clusters. The most direct test would apply CosmoMap at fiducial parameters to TNG100 or TNG300 \citep{Pillepich2018First, Nelson2019IllustrisTNG} degraded to the same voxel size, although their different mass resolution would first need to be accounted for. The CAMELS grids at $256^3$ and $512^3$ would reduce the voxel size by factors of two and four.

\paragraph*{Emissivity and lightcone model.} We use bolometric bremsstrahlung computed from voxel-averaged density and mass-weighted temperature, which neglects sub-voxel clumping, so absolute luminosities would be underestimated even with perfect predictions. In the soft band where eROSITA is most sensitive, line emission is a large fraction of the flux from gas below a few keV, so a realistic model needs a band-limited cooling function and the metallicity field. The lightcones use relative fluxes, a Gaussian PSF and a uniform background, without an instrument response or K-corrections.

\paragraph*{Input information.} Only the magnitude of the dark matter velocity is available. The velocity divergence and shear, which trace accretion shocks directly, would probably improve temperature prediction most.

\subsection{Comparison with alternative approaches}

Semi-analytic models and scaling-relation-based painting can populate haloes with integrated gas properties at negligible cost, but they do not provide the spatially resolved, morphologically diverse gas distribution needed for realistic mock images, and they ignore gas outside haloes. Our binned and tree baselines represent the simplest local versions of such mappings, and they fall well short of a 3D network, particularly at small scales and in hot gas around haloes. Generative and Lagrangian painting methods \citep{Troster2019Painting, Dai2021LDL, Horowitz2022HyPhy, Bernardini2022EMBER, Bernardini2025EMBER2, Cerardi2025XXL, Lee2026BIND} can produce sharper fields and realistic small-scale scatter, which would address the smoothing of compact structures we find, at the cost of more complex training and no direct per-voxel uncertainty. CosmoMap complements these methods with joint 3D prediction of temperature and density, explicit parameter conditioning and a benchmarked evaluation.

\subsection{Outlook}

Using CosmoMap for eROSITA-like survey modelling requires training on gravity-only inputs and on larger, higher-resolution volumes that cover the cluster mass range; converting the emissivity to counts with a band-limited cooling function that includes line emission, K-corrections, absorption and the instrument response; reproducing compact structures more faithfully; and tiled inference that keeps the normalisation statistics consistent with training. With these in place, a model that predicts a $128^3$ box in about 2\,s on a consumer GPU would allow selection functions and detection rates to be explored as functions of cosmology and feedback, which is impractical with hydrodynamical simulations alone. Predicting metallicity and pressure, and testing transfer between simulation suites with different feedback models, are further natural extensions.

\section{Conclusions}
\label{sec:conclusions}

We have presented CosmoMap, a FiLM-conditioned 3D U-Net that predicts gas temperature and electron density from dark matter density and velocity across the CAMELS IllustrisTNG parameter space. With no target statistics used at test time, it reaches $R^2 = 0.765$ for temperature and $0.934$ for electron density at $z=0$, reduces the error of gradient-boosted trees by 22--23 per cent, outperforms them at every redshift up to $z=2$, and loses little accuracy when its inputs are taken from matched gravity-only simulations. Conditioning on the simulation parameters accounts for most of the gain over simpler models, especially in low-density gas, and of the models trained with a squared-error loss, only the full conditioned network gives box luminosities that are reproducible between training seeds. Temperature is predicted best in halo cores ($R^2 = 0.88$), but the electron density of the brightest gas is biased low by $0.12\dex$. The predictions preserve the scaling relations, the shapes of surface brightness profiles, box-integrated luminosities to a median of $-0.16\dex$ (sensitive at the $0.2\dex$ level to the inference scheme unless batch normalisation is used) and most of the cosmological information in the gas. Training with a heteroscedastic likelihood gives per-voxel uncertainties at the cost of accuracy and a larger luminosity bias, and the uncertainties need rescaling as a function of predicted emissivity. In summary, $\Omega_{\mathrm{m}}$ recovered from CosmoMap's predicted gas has $\Omega_{\mathrm{m,pred}}/\Omega_{\mathrm{m,true}} = 1.01 \pm 0.11$, close to the $1.01 \pm 0.09$ obtained from the simulated gas (a partly circular test, because the network is given the parameters), and mock X-ray lightcones built from its predictions recover 89 per cent of the brightest tenth of sources, with 94 per cent of the brightest predicted sources real. The main remaining weakness is the smoothing of compact structures, which reduces the number of sources detected in mock X-ray lightcones by a factor of about three, mostly at the faint end.

\section*{Acknowledgements}

We thank the CAMELS collaboration for making their simulations publicly available.

\section*{Data Availability}

The CAMELS simulations used in this work are publicly available at \url{https://camels.readthedocs.io}. The code used to produce the results in this paper is available at \url{https://github.com/amanbh1/CosmoMap}. The trained models and derived data products will be shared on reasonable request to the authors.



\bibliographystyle{mnras}
\bibliography{references} 




\appendix

\section{Baseline details}
\label{app:baselines}

The gradient-boosted trees use up to 2000 boosting iterations with early stopping on a held-out 10 per cent of the training voxels, which stopped training after 727 and 579 iterations for temperature and density. Training on 160 instead of all 800 training simulations lowers their $R^2$ at $z=0$ by only $0.024$ for temperature and $0.008$ for density, and a run capped at 400 iterations gives the same accuracy to within $0.001$, so we do not expect a substantially stronger tree baseline from more data or iterations.

\section{Sensitivity to the inference scheme}
\label{app:inference}

Table~\ref{tab:inference} compares inference schemes for the first seed of CosmoMap, CosmoMap-$\sigma$, the small network and the unconditioned network. Voxel-level accuracy is insensitive to the scheme, but box-integrated luminosities and the unconditioned network depend on the window size through the group-normalisation statistics (Section~\ref{sec:limitations}). At a fixed window size, increasing the margin has a smaller effect, and for $64^3$ windows a margin of 16 voxels is converged. The luminosity gap between CosmoMap and CosmoMap-$\sigma$ is $0.30$--$0.39\dex$ for every scheme, and the small network is biased low by at least $0.35\dex$ for every scheme, so the conclusions of the ablation (Section~\ref{sec:ablation}) do not depend on the inference scheme. The batch-normalised retraining of CosmoMap gives the same accuracy and luminosity, to within $0.001$, for every scheme.

\begin{table*}
\centering
\caption{Test-set accuracy at $z=0$ for different inference schemes (first seed). Windows tile the $128^3$ box periodically; only the central core of each window is kept. The full-box scheme predicts the whole box in one pass with periodic padding. The scheme used throughout the paper is marked in bold.}
\label{tab:inference}
\small
\begin{tabular}{llccc}
\toprule
Model & Window (core, margin) & $R^2$ $\log T$ & $R^2$ $\log\nel$ & median $\log\frac{L_{\mathrm{pred}}}{L_{\mathrm{true}}}$ \\
\midrule
CosmoMap & $\mathbf{64^3}$ \textbf{(32, 16)} & 0.764 & 0.934 & $-0.15$ \\
CosmoMap & $96^3$ (32, 32) & 0.763 & 0.934 & $-0.03$ \\
CosmoMap & $128^3$ (64, 32) & 0.762 & 0.934 & $+0.04$ \\
CosmoMap & full box & 0.760 & 0.933 & $+0.03$ \\
U-Net, no conditioning & $\mathbf{64^3}$ \textbf{(32, 16)} & 0.646 & 0.906 & $-0.22$ \\
U-Net, no conditioning & $96^3$ (32, 32) & 0.657 & 0.902 & $-0.07$ \\
U-Net, no conditioning & full box & 0.660 & 0.853 & $-0.01$ \\
CosmoMap-$\sigma$ & $\mathbf{64^3}$ \textbf{(32, 16)} & 0.735 & 0.923 & $-0.45$ \\
CosmoMap-$\sigma$ & $96^3$ (32, 32) & 0.737 & 0.923 & $-0.34$ \\
CosmoMap-$\sigma$ & $128^3$ (64, 32) & 0.737 & 0.923 & $-0.31$ \\
CosmoMap-$\sigma$ & full box & 0.736 & 0.923 & $-0.36$ \\
Small network & $\mathbf{64^3}$ \textbf{(32, 16)} & 0.750 & 0.924 & $-0.49$ \\
Small network & $96^3$ (32, 32) & 0.747 & 0.924 & $-0.38$ \\
Small network & $128^3$ (64, 32) & 0.745 & 0.924 & $-0.35$ \\
Small network & full box & 0.736 & 0.921 & $-0.37$ \\
CosmoMap, batch norm. & every scheme above & 0.761 & 0.935 & $-0.18$ \\
\bottomrule
\end{tabular}
\end{table*}

\section{Additional figures}
\label{app:figs}

\begin{figure*}
\centering
\includegraphics[width=\textwidth]{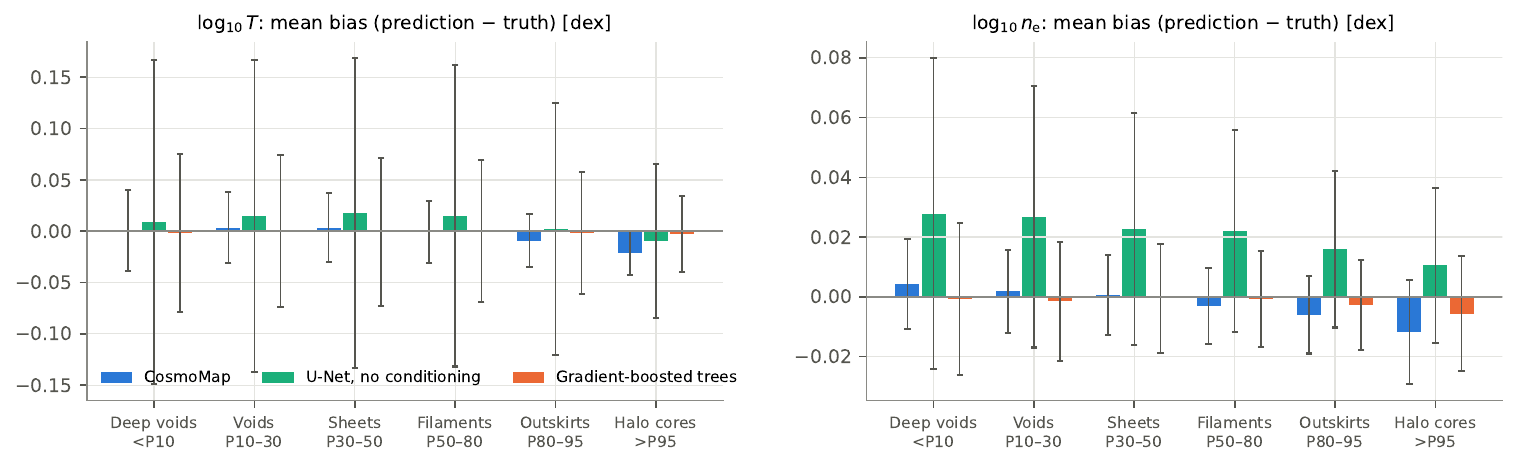}
\caption{Mean bias (prediction minus truth) of temperature (left) and electron density (right) within six ranges of dark matter density at $z=0$. Error bars show the scatter over the 100 test boxes.}
\label{fig:bias}
\end{figure*}

\begin{figure*}
\centering
\includegraphics[width=0.75\textwidth]{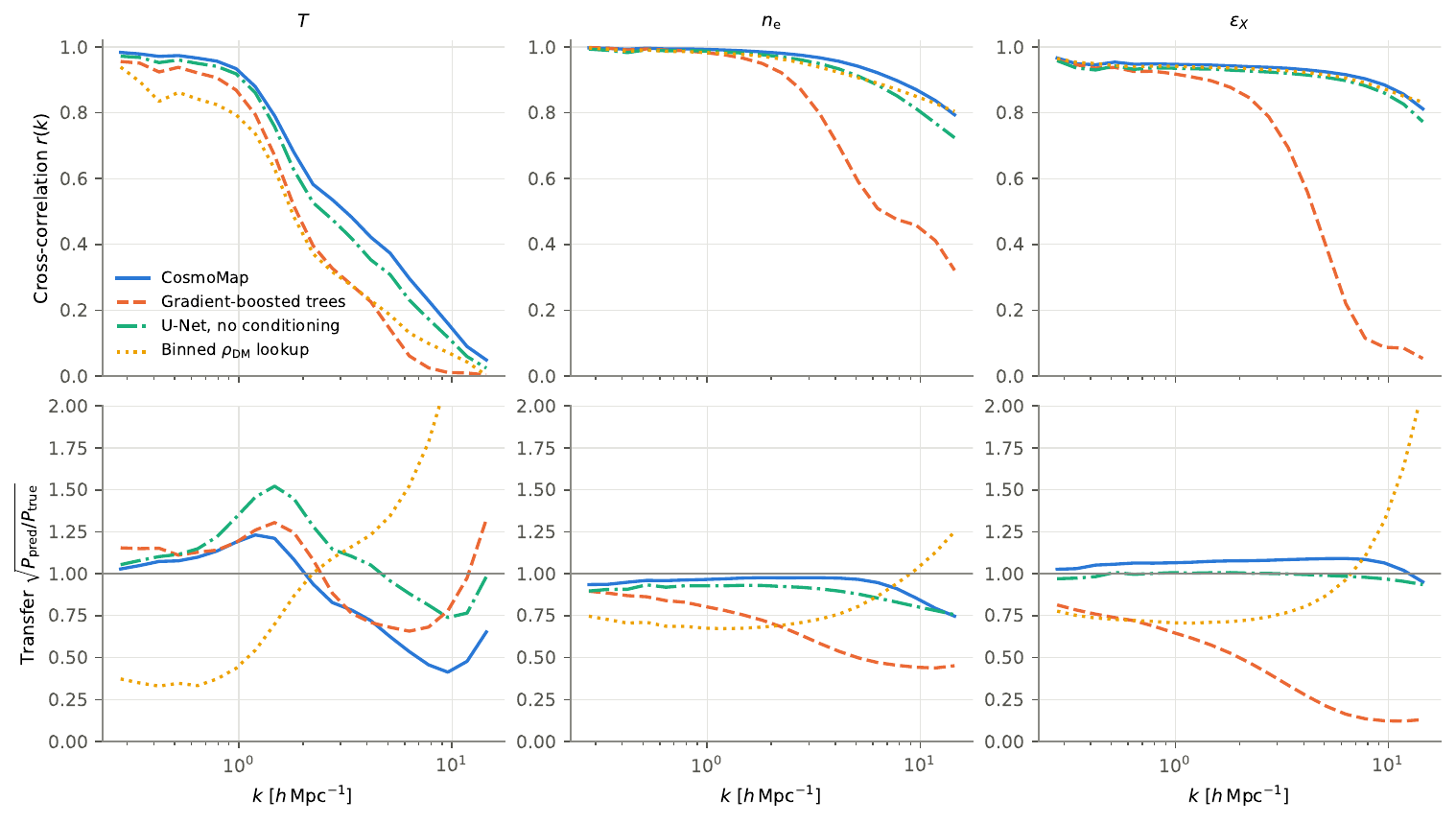}
\caption{Cross-correlation coefficient $r(k)$ (top) and transfer function $\sqrt{P_{\mathrm{pred}}/P_{\mathrm{true}}}$ (bottom) for temperature, electron density and emissivity at $z = 0$, averaged over test boxes.}
\label{fig:power}
\end{figure*}

\begin{figure*}
\centering
\includegraphics[width=\textwidth]{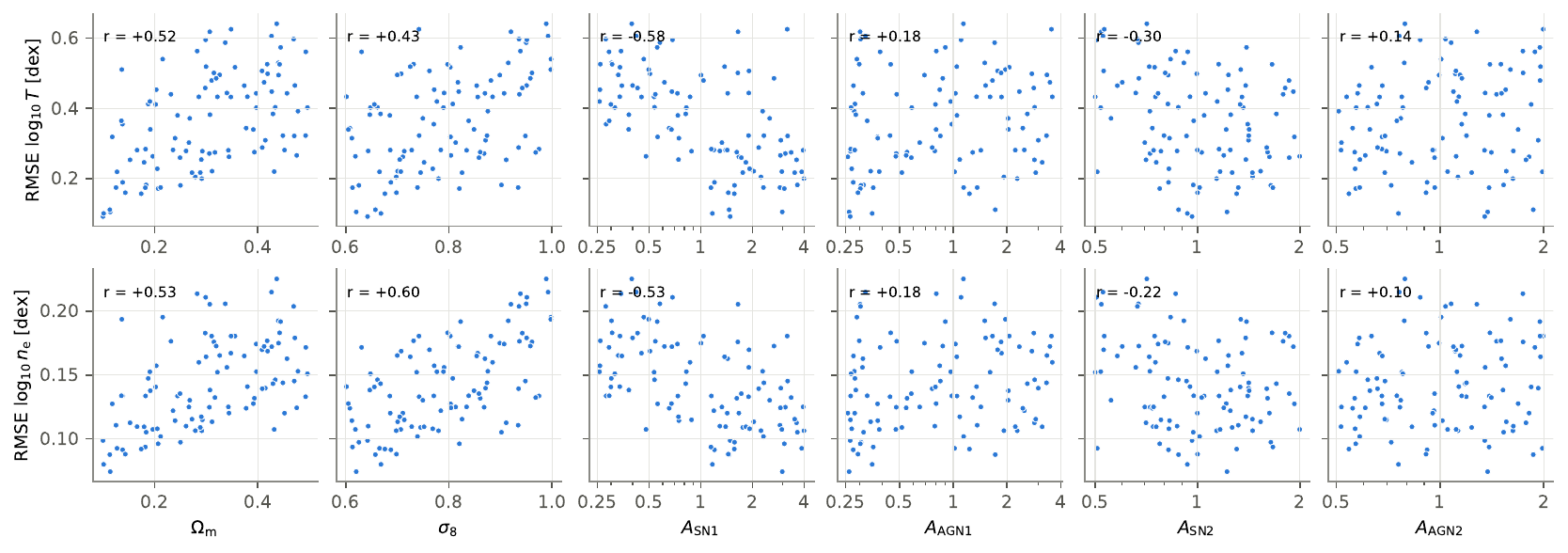}
\caption{Per-box RMSE of CosmoMap at $z=0$ against the six CAMELS parameters, with Pearson correlation coefficients (against the logarithm for the four feedback parameters).}
\label{fig:params}
\end{figure*}

\begin{figure*}
\centering
\includegraphics[width=0.75\textwidth]{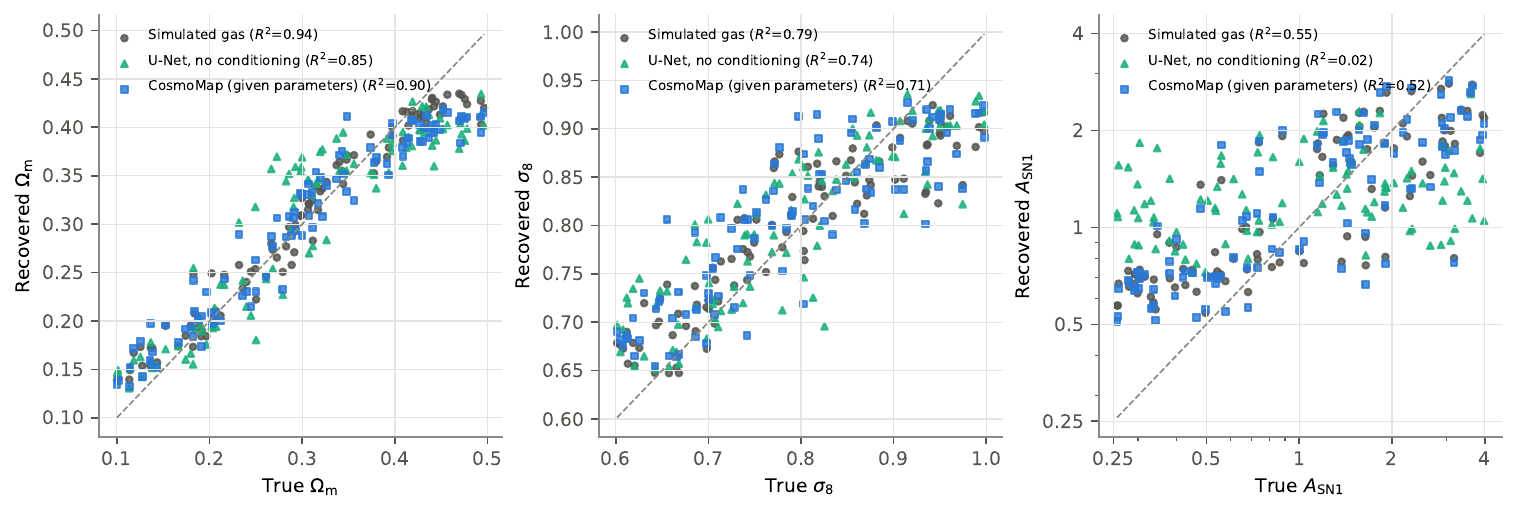}
\caption{Recovered against true $\Omega_{\mathrm{m}}$, $\sigma_8$ and $A_{\mathrm{SN1}}$ for the 100 test boxes, from summary statistics of the true gas fields, the unconditioned U-Net and CosmoMap.}
\label{fig:paramscatter}
\label{lastpage}
\end{figure*}

Figure~\ref{fig:bias} shows the systematic biases by density regime, Fig.~\ref{fig:power} the scale dependence of the predictions, Fig.~\ref{fig:params} the errors against the simulation parameters, and Fig.~\ref{fig:paramscatter} the recovered parameters for individual test boxes.

\bsp	
\end{document}